\documentstyle[12pt]{article}
\begin{document}
\title{The Nonlinear Maxwell Theory---an Outline}
\author{Artur Sowa\\
        841 Orange Street \\
        New Haven, CT 06511\footnote{The author is currently 
        with the Pegasus Imaging Corporation. This work is beyond
        the scope of his obligations there and has been performed in his
        free time. No other institution has been helpful to the author
        in conducting this research.}}
\date{March 2nd, 2001}
\maketitle
\newtheorem{th}{Theorem}
\newtheorem{prop}{Proposition}
\renewcommand{\l}{\bigtriangleup_{A}}
\begin{abstract}
The goal of this paper is to sketch a broader outline of the
mathematical structures present in the Nonlinear
Maxwell Theory in continuation of work previously presented in \cite{sowa1}, \cite{sowa2}
and \cite{sowa3}.
In particular, I display new types of both dynamic and static solutions
of the Nonlinear Maxwell Equations (NM). I point out how
the resulting theory ties to the Quantum Mechanics of Correlated
Electrons inasmuch as it provides a mesoscopic description of phenomena like 
nonresistive charge transport, static magnetic flux tubes, and charge
stripes in a way consistent with both the phenomenology and the 
microscopic principles. In addition, I point at a bunch of geometric 
structures intrinsic for the theory. On one hand, the
presence of these structures indicates that the equations at hand 
can be used as `probing tools' for purely geometric exploration of
low-dimensional manifolds.
On the other hand, global aspects of these structures are in my view 
prerequisite to incorporating (quantum)
informational features of Correlated Electron Systems within
the framework of the Nonlinear Maxwell Theory.
\end{abstract}

\section{Introduction}
The general goal of this paper is to examine broader ramifications  
of the Nonlinear Maxwell Equations (NM) as introduced by me in 1992/93
and further developed in \cite{sowa1}, \cite{sowa2}, \cite{sowa3}.
To this end, I first point out that the theory is
considerably richer than that of the classical linear Electromagnetism. 
In particular, I describe here several distinct types of both static
and dynamic solutions on a spacetime of the form $M^3\times R$.
On the technical side,
I have essentially avoided heavier analysis as the solutions are either
obtained by means of elementary calculation, or are otherwise based
on deeper analytic work described in \cite{sowa3}. One should be aware that
the possibilities opening in consequence of the introduction of these new
structures have not been fully exploited in this paper, thus postponing
many potential developments into the future.

More precisely, in the `dynamic' part of the paper I display a 
solution in the form of a charge-carrying
electromagnetic wave. It is a soliton type wave that transports 
charge with constant speed and without resistance. In addition, one 
notes existence of a specific to dimension four nonlinear Fourier type 
transform---an interesting structure whose role within the theory 
is twofold. On one hand, it can be used to find and analyze new 
solutions of the Nonlinear Maxwell Equations. On the other hand, the 
transform defines an exotic duality---a (quadratic) generalization 
of the (linear) Hodge duality. Consequences of this new duality for 
the four-geometry will be exploited in the future.

The second set of results in this paper is focused around the question of
existence and properties of static solutions. To this end, I first 
examine the situation on the Euclidean three-space. In particular, one takes
note of the occurrence of global structures in the form of
magnetic flux tubes as well as the so-called
charge stripes. It is interesting from the point of view of geometry that
these objects exist in general only on three-manifolds whose 
fundamental group is not finite. This is tied to the geometric fact that 
the nonlinear gauge theory at hand induces an additional structure on 
$M^3$---namely a taut codimension one foliation. These global aspects
of static solutions prompt an assumption of topological point of view.
Accordingly, I sketch
the possibility of constructing `nonlinear cohomology' that would
account for a sort of `flux tube' invariant of a three manifold.
The discussion here is based on two particular examples that I feel provide
an optimal illustration of the underlying concept.

The Nonlinear Maxwell Equations, cf. (\ref{syst0}-\ref{syst2}) below,
involve a vector potential that encodes the electric and magnetic
fields in the usual way as well as an additional scalar $f$. The 
function $f$ contains information, extractable in a certain simple
canonical way, about the local value of the {\em filling factor} (also
known as the filling fraction). (The filling factor is defined as 
the number of quanta of the magnetic field per electron charge in the first
Landau level. It is then natural and effective to think of the electrons
as forming in conjunction with the corresponding magnetic flux quanta
composite particles---either bosons or fermions, or Laughlin particles
depending on the actual value of the filling factor.)
It is thus postulated that the filling fraction---typically an input
of a microscopic theory that is always assumed constant microlocally---is 
allowed to slowly vary in the coarser scale. In fact, it was shown
in \cite{sowa3} that NM predict occurrence of phase changes that lead
to formation of vortices in $f$, and {\em a fortiori} in the magnetic 
field. This picture conforms with the well known analogy between the Quantum
Hall Effects and the High-$T_c$ Superconductivity. An inquisitive reader
might now point at the following seeming conundrum. 
The physical interpretation of $f$ as a filling factor requires 
the presence of two-dimensional geometric structures that endow 
us with a possibility of including the lowest Landau level in the 
basic dictionary. Thus, it may appear a priori puzzling, how we are 
going to retain this interpretation of $f$ in three or four spatial 
dimensions? The answer is provided by the intrinsic structure of the 
NM themselves. On one hand, it is shown below that the filling factor 
variable may be completely factored out of the equations when viewed
in the complete four dimensions of spacetime. Needless to say, if one 
attempted to analyze such $f$-free form in two dimensions the $f$ 
variable would reemerge without change as it is there encoded in the 
magnetic field $B =b/f$, $b=\mbox{const}$.
On the other hand, one notes that a remnant or a generalization of 
the filling factor interpretation carries over to three dimensions.
Namely, the NM in three dimensions imply existence of a codimension 
one foliation of the three-space associated with the static solutions.
Moreover, one notes here that the NM do not {a priori} introduce any 
restrictions as to the type of the resulting foliation---in fact any regular 
foliation and even foliations with 
singularities introduced by degenerating leaves are admitted by the 
equations. However, as already mentioned above the existence of solutions of 
a special type, namely the flux-tube type, implies 
geometrical restrictions on the foliation and topological restrictions
on the three-manifold. Indeed, in this case foliation must be taut.
It also seems reasonable to expect that the composite 
particle interpretation remains valid in this setting and the number 
of participating electrons in each leaf is again determined by $f$, 
virtually leading to the notion of an {\em effective} Landau level.

In the last words of this section I would like to admit that, the subject 
matter at hand being both new and inherently interdisciplinary as well
as by way of my own background and limitations, it is 
not always easy to pick the optimal terminology. Realizing I will
unavoidably fail to satisfy in this respect one group of readers 
or another, I can only ask the readership to be as tolerant as they 
can afford and hope that in the end substance will triumph over form.

\section{Nonlinear Maxwell Equations in Spacetime}

In what follows, in order to get around rather tedious algebra while
not compromising
our understanding of what is essentially involved, I present a shortcut style
exposition of the necessary calculations. I believe that readers who are
well familiar with differential geometry will find it easy to
reinterpret this calculation in its natural invariant setting, while
those who are less familiar with the abstract setting may in fact appreciate
its absence here.

Consider the following system of equations---the Nonlinear Maxwell
Equations (NM) in the form in which they have appeared in my previous papers.
\begin{equation}
 dF_{A}=0
\label{syst0}
\end{equation}
\begin{equation}
 \delta (fF_{A})=0
\label{syst1}
\end{equation}
\begin{equation}
 \Box f +a|F_{A}|^{2}f=\nu f.
\label{syst2}
\end{equation}
where $f$ is a real valued function and $A$ is the electromagnetic 
vector potential, so that the corresponding electromagnetic field 
is $ F_A = dA$. Here, $a>0$ is a physical constant with
unit $\left[\mbox{Tesla}^{-2}\mbox{m}^{-2}\right]$;
I will not discuss the precise physical
interpretation of $a^2$ in this paper.
Further, $d$ is the exterior derivative and $\delta =
\star d\star$ its adjoint. Here, it is assumed that the Hodge star $\star$ 
and the D'Alembertian $\Box$ are induced by the Lorentzian metric tensor
on a spacetime of $3+1$ dimensions.
Let me point out that assuming $f=\mbox{const}$ and dragging it to zero one 
recovers the classical Maxwell equations. In this sense, all phenomena of
the classical electromagnetism are included in the present model.

Now the goal is to better understand the essential ingredients of the NM in terms
of the classical field variables.
To this end, let us say the spacetime is in fact the flat Minkowski 
space (with the speed of light $1$) so that in particular one can
identify coefficients of the electric field $\vec{E}$ and the magnetic
field $\vec{B}$  with the coefficients of the curvature tensor
$F_A$ by the formula
\begin{equation}
F_A = B_1dy\wedge dz +B_2 dz\wedge dx +B_3 dx\wedge dy +
        E_1dx\wedge dt +E_2 dy\wedge dt +E_3 dz\wedge dt.
\label{FA}
\end{equation}
For the sake of our discussion below it is good to keep in mind the well 
known fact that the components of $F_A$ are not Lorentz invariant. 
This property leads one to the derivation of the Lorentz force, so that the 
latter one is logically
independent of the particular form of a gauge theory formulated in terms 
of $F_A$. In other words, the Lorentz force remains unchanged and valid 
as one attempts to modify the field equations. With this understood,
let us continue the discussion of equations (\ref{syst0}-\ref{syst2}).

It would be rather straightforward to rewrite equations
(\ref{syst0}-\ref{syst2}) in the anticipated Maxwellian form
by following the usual procedures for translating (\ref{syst1})
into the \'Ampere and Gauss laws after having replaced
$\vec{E}$ by $f\vec{E}$ and $\vec{B}$ by $f\vec{B}$.
In fact, this would lead to an {\em ad hoc} interpretation of $f$ as a
material constant---a route taken in our older, one might say {\em naive},
paper \cite{sowa1}. However, this form of the system offers little insight
as to the more essential implications of the NM, and one needs
to find a less obvious reformulation.

One notes that equation (\ref{syst1}) may be equivalently written in the form 
\begin{equation}
f\delta F_{A}=F_A(\nabla_{3+1} f,. )
\label{twoprime}
\end{equation}
where $\nabla_{3+1}$ stands for the gradient in spacetime.
For a reason that will become clear later,
one identifies $F_A$ with a skew-symmetric matrix in a standard way
\[
F = \left(\begin{array}{rrrr}
           0 & -B_3 & B_2 & -E_1 \\
         B_3 & 0   & -B_1 & -E_2 \\
        -B_2 & B_1 & 0   & -E_3 \\
         E_1 & E_2 & E_3 & 0 
       \end{array}\right).
\]
It is important to note that by the miraculous property
of skew-symmetric matrices in four dimensions,
$\det F = \vec{E}\cdot\vec{B}$ and
\[ 
F^{-1} = \frac{1}{\vec{E}\cdot\vec{B}}
 \left(\begin{array}{rrrr}
           0  & E_3 & -E_2 & B_1 \\
         -E_3 &   0 &  E_1 & B_2 \\
          E_2 &-E_1 &  0   & B_3 \\
         -B_1 &-B_2 & -B_3 & 0 
       \end{array}\right)
       =  \frac{1}{\vec{E}\cdot\vec{B}} \hat{F}.
\]
(I emphasize that $\hat F$ is not the matrix
corresponding to the Hodge-dual of $F_A$ in the given metric with 
signature $(+++-)$.)
On the other hand, representing both the $1$-forms and vectors 
as columns so that in particular
\[
\nabla_{3+1} f = \left[\begin{array}{r}
           f_x \\
           f_y \\
           f_z \\
          -f_t 
       \end{array}\right] \mbox{ and }
\delta F_{A}= \left[\begin{array}{r}
             E_{1,t}+B_{2,z}-B_{3,y}\\
             E_{2,t}+B_{3,x}-B_{1,z}\\
             E_{3,t}+B_{1,y}-B_{2,x}\\
             E_{1,x}+E_{2,y}+E_{3,z}
       \end{array}\right],
\]
one checks directly that
\[
 F_A(\nabla_{3+1} f,. ) = F \nabla_{3+1} f.
\]
This enables us to rewrite equation (\ref{twoprime}) in the form
\begin{equation}
  F^{-1}\delta F_{A}=\nabla_{3+1} \ln (f).
\label{twoprimevec}
\end{equation}
It is perhaps worthwhile to realize that in this context $F$ is a 
fiberwise-linear mapping from the tangent bundle to the cotangent bundle.
Here one assumes $f>0$ a.e. This conforms with the principle that one will be
consistently looking for {\em strong} solutions so that in particular $f$ may always
be replaced with $|f|$ in (\ref{syst0}-\ref{syst2}). 
Next one recalls that on one hand the first part of the NM (\ref{syst0}) 
is identical 
with the analogous part of the classical Maxwell equations and it 
encodes the Faraday's law of magnetic induction and the fact that 
there are no spatially extensive magnetic charges. This gives the 
first four 
(scalar) equations below, namely (\ref{first}) and (\ref{second}). 
On the other hand, by direct multiplication 
and regrouping in (\ref{twoprimevec}), one obtains four further scalar 
equations that happen to radically modify the \'{A}mpere law.
Written in the familiar three-space vector notation, the NM assume the form
\begin{equation}
\frac{\partial\vec{B}}{\partial t}+\nabla\times\vec{E}=0
\label{first}
\end{equation}
\begin{equation}
\nabla\cdot\vec{B} =0
\label{second}
\end{equation}
\begin{equation}
(\frac{\partial\vec{E}}{\partial t}-\nabla\times\vec{B})\times\vec{E}
+(\nabla\cdot\vec{E})\vec{B} = -(\vec{E}\cdot\vec{B})\nabla\ln f
\label{third}
\end{equation}
\begin{equation}
\label{fourth}
(\frac{\partial\vec{E}}{\partial t}-\nabla\times\vec{B})\cdot\vec{B}
= -(\vec{E}\cdot\vec{B})\frac{\partial}{\partial t}\ln f
\end{equation}
\begin{equation}
(\frac{\partial ^2}{\partial t^2} - \bigtriangleup )
f + a(|\vec{B}|^2 - |\vec{E}|^2)f = \nu f,\vspace{1.5cm}
\label{last}
\end{equation}
provided $\vec{E}\cdot\vec{B} \neq 0$. In fact, also the case when
$\vec{E}\cdot\vec{B} = 0$ is worth our attention and will be discussed
below. As much as one should avoid indulging in formal manipulations
of formulas, here the advantage of having the equations rewritten
in several equivalent forms is that they all 
lead to the discovery of new types of solutions, the existence of which 
would be otherwise obscured by notation. This will become more evident in the
following sections.

As already mentioned in the introduction, 
I postulate the following physical interpretation: $f$ is 
the spatially varying filling factor---a notion central to the 
modern composite-particle theories. In fact, the canonical 
microscopic-theory  interpretation
of the filling factor
is valid in two spatial dimensions only, in which case it signifies the 
ratio of the number of quanta of the ambient magnetic field to 
the number of electrons in the first Landau level, cf. \cite{Laugh1},
\cite{pr}.
 Moreover,
the microscopic theory offers no hints as to the existence and relevance 
of an analogous notion in the three-space. 
A description of the interaction of the electromagnetic field with
fermions in the first Landau level provided by the equations above 
is valid in the mesoscopic scale. Here, as one `zooms out' from the 
microscopic scale, the filling factor is neither a rational number 
nor is it a constant anymore. In fact, as it has been communicated 
in previous papers the spatially varying filling factor 
may assume the form of a vortex lattice, cf. \cite{sowa3}.
For the time being, this point
of view is validated by the well known analogy between the Quantum Hall
Effect and the High Temperature Superconductivity and it awaits 
experimental confirmation.
Moreover, the NM extend
the notion of the filling factor to three spatial dimensions. However,
as we will see below, the presence of the filling factor introduces an 
especially interesting 
modification of the laws of Electromagnetism only if the three-space comes 
equipped with a codimension one foliation. This latter fact makes it possible to 
talk about Landau levels in a certain sense, anyhow.
Finally, $f$ can be completely eliminated from the NM in $3+1$ dimensions. 
(In general, this 
requires that the first cohomology group of the spacetime vanishes.) In that case 
the NM can be written in the $f$-free form
\[
dF_A=0
\]
\begin{equation}
d\left(F^{-1}\delta F_{A}\right)^\# = 0
\label{integr}
\end{equation}
\begin{equation}
 \delta\left(F^{-1}\delta F_{A}\right)^\#
 -\left|F^{-1}\delta F_{A}\right|^2 + 
a|F_A|^2 - \nu =0,
\label{madeq}
\end{equation}
where $\#$ is the isomorphism of the tangent and the cotangent bundles 
given by the metric.
Indeed, under the assumption of vanishing first de-Rham cohomology,
equation (\ref{twoprimevec})
is equivalent to its integrability condition (\ref{integr}).
Moreover, since the last scalar equation of the system can be written 
in terms of $d\ln f$ in the form
\[
\delta d\ln f -|d\ln f|^2 + a|F_A|^2 - \nu =0,
\]
equation (\ref{twoprimevec}) also implies (\ref{madeq}).
Computation of the symbol shows that the system obtained in 
this way is non-hyperbolic---in fact its degeneracy is of higher order.
Thus, this form of the NM appears impractical for any mathematical work,
and an introduction of the dimensionless scalar $f$ is necessary also 
from the point of view of analysis. Nevertheless, as indicated in the Introduction
and the discussion above, physical implications
of the existence of an $f$-free form of the NM are important.

\section{Geometry Behind the Equations}

The geometrical arena of the Maxwell equations consists of 
a spacetime, say $N$, and a principal $U(1)$-bundle, say $P$,
stack up above $N$.
In addition, it seems any description of the interaction of 
the electromagnetic field with fermions requires, at least 
within this framework, a principal connection, i.e. a smooth (at least 
a.e.) distribution of horizontal planes that is invariant with respect 
to the circle action. This distribution can be written as $\ker{A} = 
\ker{fA}$ for $f\neq 0$. In addition, if $U(1)$ is to remain the elemental 
symmetry group of Electromagnetism, then $ f$ must be constant along the 
fibers so that it effectively descends to a function on $N$. In particular,
within this dictionary one can construct a Kaluza-Klein metric on $P$,
which is given by
\[ \mu_A(X,Y) = g(\pi_*X,\pi_*Y)+aA(X)A(Y),
\]
where the unit of $a>0$ must be 
$\left[\mbox{Tesla}^{-2}\mbox{meter}^{-2}\right]$ if the unit of length on $P$
is to be $\left[\mbox{meter}\right]$ and the unit of $F_A =dA$ is to remain, say, 
$\left[\mbox{Tesla}\right]$.
Let us say the corresponding Laplace-Beltrami operator on forms is then
$\bigtriangleup_{\mu_A}=\bigtriangleup_A$. Calculation shows that the 
condition
\[\bigtriangleup_A (fA) =\nu fA
\]
is equivalent to the system of equations (\ref{syst0}--\ref{syst2}), cf. \cite{sowa1}.

\section{Exotic Duality}
For the sake of discussion in this section, consider the NM on
either a Lorentzian or a Riemannian four-manifold
as the metric signature plays a secondary role. In particular,
it is preferable to replace the $\Box$-notation with the 
$\bigtriangleup$-notation. Assume for the sake of simplicity
that the second cohomology group of the manifold is trivial.
Omitting the constant $a$, write 
the system one more time in the form
\begin{equation}
\label{A1}
\delta (f dA) = 0
\end{equation}
\begin{equation}
 -\bigtriangleup f +|dA|^{2}f=\nu f.
\label{A2}
\end{equation}
Since (\ref{A1}) implies $d(f\star dA) =0$, one has
$f\star dA = d\tilde{A}$ so that
\begin{equation}
\label{transform}
dA = \pm 1/f\star d\tilde{A}
\end{equation}
and the new form $\tilde{A}$ satisfies a {\em dual} system of equations
\begin{equation}
\label{tildA1}
\delta (\frac{1}{f} d\tilde{A}) = 0
\end{equation}
\begin{equation}
 -\bigtriangleup f +|d\tilde{A}|^{2}\frac{1}{f}=\nu f.
\label{tildA2}
\end{equation}
This is a functional transform reminiscent of the Fourier or 
the Backlund transforms, notwithstanding the fact that all 
transforms are somewhat reminiscent of one another.
In particular, the resulting dualistic perspective
has the expected property that trivial solutions of one of the systems
lead to more complex solutions of the dual system. To illustrate the 
idea, let me now present a few examples of dual solutions on $R^4$ with
either the Euclidean or the Minkowski metric as specified in the discussion.

{\em Example 1.} Let the metric be Euclidean and
 take $dA = Edz\wedge dt$, $E=\mbox{const}$,
and $f=f(x,y)$. Equation
(\ref{A1}) is automatically satisfied and (\ref{A2}) assumes the form
$-f_{xx} - f_{yy} = (\nu - E^2)f$ so that 
$f=\cos{(k_1x+k_2y+\alpha )}$ for 
$k_1^2+k_2^2 = \nu - E^2$ solves the problem. Now
$d\tilde{A} = \pm f(x,y) dx\wedge dy$ and it staisfies equations
(\ref{tildA1}-\ref{tildA2}).

{\em Example 2.} Departing for a while from the assumption 
of vanishing second cohomology, let us reinterpret the previous example on a 
four-torus assuming periodicity of coordinates $(x,y,z,t)$
with period $2\pi$. Note 
that the first bundle is necessarily nontrivial as the cohomology 
class $[dA]\neq 0$. Let us allow the function $f$ drop its dependence 
on $y$ so that, say, $f = \cos x$, provided the `right choice' of $\nu$
has been made.
Now, $d\tilde{A} = d(\sin x dy)$ is an exact form so the second
bundle is topologically trivial.

{\em Example 3.} Consider $dA = Bdx\wedge dy$ and $f=f(z,t)$ so that
(\ref{A1}) is satisfied. Let us now look at the metric with
signature $(+++-)$ so that (\ref{A2}) means $f_{tt} - f_{zz} = 
(\nu - B^2)f$. The general solution of this equation is a standing 
wave with variable amplitude. This pattern is inherited by 
$d\tilde{A} = f(z,t)dz\wedge dt$ (up to the sign again) which satisfies
(\ref{tildA1}-\ref{tildA2}).

{\em Example 4.} Let us for a change begin on the other side and take, say,
$d\tilde{A} = edz\wedge dt$ and $f=f(x,y)$. Again, the first equation
(\ref{A1}) is automatically satisfied while (\ref{A2}) becomes 
$-f_{xx} - f_{yy} + e^2/f = \nu f$. As explained in
\cite{sowa3} (see also remarks at the end of section \ref{static} below)
apart from the trivial constant solution, this problem also has a solution
in the form of a vortex lattice. In the latter case
$dA = f(x,y) dx\wedge dy$ satisfies (\ref{A1}-\ref{A2}) and represents
static magnetic flux tubes.

I emphasize that only the vector potential $A$ and the filling fraction 
variable $f$ that appear in the first set of equations have 
physical interpretation. Reassuringly, the presence of 
a nontrivial $f$ in examples {\em 1} and {\em 3} did not contribute anything 
unexpectedly strange to the constant electric and magnetic fields in these
examples, while it `introduced' flux tubes in example {\em 4}.   
Although one could consider similar interpretation of the transformed vector 
potential $\tilde {A}$, just as one can for any $U(1)$-connection, 
I feel this is uncalled for and would probably be unjustifiable 
at this point. Nevertheless, the existence of the transform is a 
remarkable fact whose possible applications to four-manifolds will 
be explored more thoroughly in the future. In a way, this new duality
is a generalization of the regular Hodge-star duality that may be 
compared to the projective generalization of the Euclidean reflection.
This analogy may be justified in the following way.
Projective duality is induced by a fixed quadratic form.
What is the NM analog of that object?
Introduce notation $\varphi = \ln f$. A
direct calculation shows that (\ref{A1}-\ref{A2}) may be written in the 
form of a system of quadratic equations
\begin{equation}
\label{AA1}
 \delta dA + \star (d\varphi\wedge \star dA) = 0
\end{equation}
\begin{equation}
\label{AA2} 
-\bigtriangleup\varphi -|d\varphi |^2 + |dA|^2 -\nu = 0.
\end{equation}
This form of the equation has one other advantage. Suppose one has
found a solution $(A, \varphi )$ of (\ref{AA1}-\ref{AA2}). One can now 
use gauge invariance of the equations in the following way. Let 
$\chi$ be a solution of the equation
\[
\delta d \chi = -\delta A.
\]
The existence of a solution $\chi$ follows from the Fredholm alternative
when the metric is positive definite, and it amounts to solving a linear
wave equation in a Lorentzian metric. One can now replace $A$ with 
$A+d\chi$ (and denote the resulting form by $A$ again). 
In the new gauge $\delta A =0$, so that $A$ in fact satisfies 
\begin{equation}
\label{AAA1}
 \bigtriangleup A + \star (d\varphi\wedge \star dA) = 0.
\end{equation}
The system that consists of (\ref{AAA1}) and (\ref{AA2}) is either
quasilinear elliptic or hyperbolic, depending on the metric. 
Solving the latter system may 
not be helpful at all in finding solutions of the original 
(\ref{AA1}-\ref{AA2}), since one cannot guarantee that a solution 
satisfies the Lorentz gauge condition $\delta A =0$. However, 
solutions of (\ref{AA1}-\ref{AA2}) {\em a fortiori} satisfy
(\ref{AAA1}) and (\ref{AA2}) so that in particular they will obey
all a priori estimates on the solutions of, say, quasilinear
hyperbolic systems.
In particular, this point of view may justify the claim
that the phenomena described in this paper shed some light on 
the complex nature of quasilinear systems of PDE of certain types in general. 
        
\section{Charge Transport and Charge Stripes}
\label{charge_wave}

I will now take full advantage of the (\ref{first}-\ref{last}) form of the
NM. In analogy to the electromagnetic 
wave in vacuum, that one recalls is counted among the solutions of this
system, one wants to look for a solution with $\vec{E}\cdot\vec{B} = 0$. 
In the end I will check that the new solution of (\ref{first}-\ref{last}) 
in fact satisfies
(\ref{syst0}-\ref{syst2}) which is not a priori guarantied.
Make an Ansatz 
\begin{equation}
\label{EB}
\vec{B}=B_1\frac{\partial}{\partial x}
+ B_2\frac{\partial}{\partial y},
\qquad
\vec{E}=e\left(-B_2\frac{\partial}{\partial x}
+ B_1\frac{\partial}{\partial y}\right), 
\end{equation}
where $e$, $B_1$ and $B_2$ are a priori functions of $(x,y,z,t)$ that are 
smooth a.e. and neither one of them vanishes identically. 
As an immediate consequence, one obtains that (\ref{first}) and
(\ref{second}) are equivalent to
\begin{equation}
\label{helper1}
B_{1,t} = (eB_1)_{,z}
\end{equation}
\begin{equation}
\label{helper2}
B_{2,t} = (eB_2)_{,z}
\end{equation}
\begin{equation}
(eB_1)_{,x} +(eB_2)_{,y} = 0
\end{equation}
\begin{equation}
B_{1,x} + B_{2,y} = 0
\end{equation}
which implies
\begin{equation}
e_{,x}B_1 +e_{,y}B_2 = 0.
\end{equation}
On the other hand, (\ref{third}) and (\ref{fourth}) are equivalent to 
\begin{equation}
(B_{2,x} -B_{1,y})eB_1 = (eB_2)_{,x}B_1 - (eB_1)_{,y}B_1 
\end{equation}
\begin{equation}
(B_{2,x} -B_{1,y})eB_2 = (eB_2)_{,x}B_2 - (eB_1)_{,y}B_2 
\end{equation}
\begin{equation}
\label{longy}
\left(-(eB_2)_{,t} +B_{2,z}\right)B_1 + \left((eB_1)_{,t} - B_{1,z}\right)B_2=0. 
\end{equation}
Equations (\ref{helper1}), (\ref{helper2}) and
(\ref{longy}) imply that $e$ is in fact constant 
\begin{equation}
\label{thee}
 e = \pm 1.
\end{equation}
Using (\ref{helper1}) and (\ref{helper2}) again, one obtains
\[
B_1 = B_1(x, y, t+ez), \qquad B_2 = B_2(x, y ,t+ez).
\]
In particular, $ \vec{B}$ and $\vec{E}$ are not compactly supported.
At this point, the only condition left {a priori}
unfulfilled is the vanishing divergence condition.
Thus, all equations (\ref{helper1}-\ref{longy}) above are satisfied iff
there is a function $\psi = \psi (x, y, t+ez)$ such that 
\begin{equation}
\label{thebs}
B_1 = -\psi_y(x, y, t+ez),\qquad B_2 = \psi_x(x, y, t+ez).
\end{equation}
Defining the electric and magnetic fields by (\ref{EB})  with $e=\pm 1$ ,
so that in particular $|\vec{E}| =|\vec{B}|$, and choosing $f$ that satisfies 
the linear wave equation (\ref{last}), one obtains a solution of
(\ref{first}-\ref{last}) .

However, physical solutions must in addition satisfy the {\em a priori}
more restrictive system
(\ref{syst0}-\ref{syst2}). Consider $F_A$ as given in (\ref{FA}). 
Equation (\ref{syst0}) is satisfied automatically since it is equivalent to 
(\ref{first}-\ref{second}). On the other hand, (\ref{syst1}) becomes
\begin{equation}
\label{syst1_1}
(fB_2)_{,x} - (fB_1)_{,y} = 0
\end{equation}
\begin{equation}
\label{syst1_2}
(feB_2)_{,t} - (fB_2)_{,z} = 0
\end{equation}
\begin{equation}
\label{syst1_3}
-(feB_1)_{,t} + (fB_1)_{,z} = 0
\end{equation}
\begin{equation}
\label{syst1_4}
-(fB_1)_{,y} + (fB_2)_{,x} = 0
\end{equation}
Now, (\ref{syst1_2}) and (\ref{syst1_3}) imply via (\ref{thebs}) that
\[
 f = f(x, y, t+ez).
\]
In particular, $f_{,tt} - f_{,zz} =0$. Thus, (\ref{syst0}-\ref{syst2})
has been reduced to the following system of two equations:
\begin{equation}
\label{smallsyst1}
-f_{,xx} - f_{,yy} = \nu f 
\end{equation} 
\begin{equation}
\label{smallsyst2}
(f\psi_{,x})_{,x} + (f\psi_{,y})_{,y} = 0.
\end{equation}
The first equation above admits three types of classical solutions. Namely,
\begin{equation}
f = \left\{ \begin{array}{lll}
    A(t+ez)\ln{(x^2+y^2)}   &\quad \nu =0 \\
    A(t+ez)\cos {(k_1x+k_2y+\alpha(t+ez))} &\quad \nu = k_1^2+k_2^2 \\
A(t+ez)\exp {(k_1x+k_2y)} &\quad \nu = -k_1^2-k_2^2 .
\end{array}\right.
\label{thef}
\end{equation} 
Observe that each solution effectively depends on one harmonic variable in the
$(x,y)$-domain---either, 
$u = k_1x+k_2y$ or $u=\ln{r^2} =\ln{(x^2+y^2)}$. Thus, equation (\ref{smallsyst2})
is satisfied if 
\[
\psi_u = C(t+ez)/f(u,t+ez),
\]
for an arbitrary function $C$ of one variable. Therefore, in view of (\ref{thebs})
 one obtains three types of solutions (redefining $C$)
\begin{equation}
\label{theBs}
 [B_1,B_2] = \left\{\begin{array}{lll}
C(t+ez)/(r^2\ln{r^2})[-y,x] \\
 C(t+ez)\sec{(k_1x+k_2y +\alpha)}[-k_2,k_1]\\
 C(t+ez)\exp{(-k_1x-k_2y )}[-k_2,k_1]
\end{array}\right.
\end{equation}
in correspondence with (\ref{thef}).
  Since one is looking for strong solutions,
one has the freedom to cut off pieces of the classical solutions (by restricting 
the domain) and to put them back together. In this way, one obtains solutions that 
are either continuous
or have jump discontinuities but may be guarantied to remain bounded. Last but not least,
it is physically correct
to interpret the divergence of the electric field as charge $\rho$ and 
$-\frac{\partial}{\partial t}\vec{E}+\nabla\times\vec{B}$ as the electric current.
One checks that for solutions as above the $(x,y)$-component of current vanishes while
the $z$-component $j$ is equal to $-e\rho$. More precisely, 
one obtains that piecewise
\begin{equation}
 e\rho =-j= \left\{\begin{array}{lll}
4C(t+ez)/(r^2\ln  ^2r^2) \\
\nu C(t+ez)\sec{(k_1x+k_2y+\alpha)}\tan{(k_1x+k_2y+\alpha)}\\
-\nu C(t+ez) \exp{(-k_1x-k_2y)}
\end{array}\right. 
\end{equation}
in correspondence with (\ref{thef}) and (\ref{theBs}).
In addition to the piecewise smooth distribution of charge,  
one should include charge concentrated on 
singular surfaces where the electric field has jump discontinuities 
as indicated by the distributional derivative 
$\nabla \cdot \vec{E}$. Therefore, charge is transported along the z-axis 
with the speed $e=\pm 1$ and without resistance as the vector of current 
is perpendicular to the electric field. Charge is mostly 
concentrated along {\em charge stripes} where the electric and magnetic 
fields have singularities.  
The net current depends on the particular choice of a (strong) 
solution. Of course, the theory does not tell us how to solve 
the practical problem of electronics---namely, how to create conditions 
for a particular function $C = C(t+ez)$, constant $\nu$  and a desired 
mosaic of singularities to actually occur in a physical system.

\section{Static Solutions and Magnetic Flux Tubes}
\label{static}
The classical Maxwell equations admit static solutions
of two types only: the uniform field solutions, and the unit charge
or monolpole-type solutions, as well as superpositions of these 
fundamental types of solutions. As we will see below, the nonlinear theory 
encompasses a larger realm including the magnetic-flux-tube type and the 
charge-stripe type solutions. These additional configurations require
nonlinearity
and cannot be superposed, which gives them more rigidity. In the next section
we will see what can be said about the variety of such solutions, while in
this section I will only display a single example of this type.
Apart from the applicable goal, the idea is to present an example
that possesses all the 
essential features of the general class of solutions 
yet the required calculation is free of more subtle geometric technicalities.

Time-independent solutions of the NM posses physical interpretation
only if they satisfy the equations in the classical sense almost everywhere.
Assuming that all fields are independent of time (\ref{first}-\ref{last}) 
takes on the form
\begin{equation} 
\nabla\times\vec{E}=0
\label{first_st}
\end{equation}
\begin{equation}
\nabla\cdot\vec{B} =0
\label{second_st}
\end{equation}
\begin{equation}
-(\nabla\times\vec{B})\times\vec{E}
+(\nabla\cdot\vec{E})\vec{B} = -(\vec{E}\cdot\vec{B})\nabla\ln f
\label{third_st}
\end{equation}
\begin{equation}
(\nabla\times\vec{B})\cdot\vec{B} = 0
\label{fourth_st}
\end{equation}
\begin{equation}
- \bigtriangleup f + a(|\vec{B}|^2 - |\vec{E}|^2)f = \nu f,
\label{last_st}
\end{equation}
under the assuption that $\vec{E}\cdot\vec{B}\neq 0$ a.e.
Adopt an Ansatz that the integral surfaces of the planes 
perpendicular to the field $\vec{B}$ are flat, say,
\[
\vec{B} = b(x,y)\frac{\partial}{\partial z}.
\]
One easily checks that equations (\ref{second_st}) and (\ref{fourth_st}) are satisfied.
Assume in addition that the electric field is potential, i.e.
\[
\vec{E} = \nabla \psi(x,y,z), \mbox{ where } \psi_z \neq 0 \mbox{ a.e.} 
\]
so that (\ref{first_st}) is satisfied. Remembering notation $\varphi = \ln f$,
one calculates directly that
\[
(\nabla\times\vec{B})\times\vec{E} = 
-\psi_zb_x\frac{\partial}{\partial x}
-\psi_zb_y\frac{\partial}{\partial y}
+(\psi_xb_x+\psi_yb_y)\frac{\partial}{\partial z},
\]
while
\[
(\nabla\cdot\vec{E})\vec{B} =
\bigtriangleup\psi b\frac{\partial}{\partial z},
\]
and
\[
(\vec{E}\cdot\vec{B})\nabla\varphi.
\]
Thus, equation (\ref{third_st}) is equivalent to the following system
of three equations
\[
\psi_z(b\varphi_x +b_x)=0
\]
\[
\psi_z(b\varphi_y +b_y)=0
\]
\[
b\bigtriangleup\psi -\psi_xb_x-\psi_yb_y-b\psi_z\varphi_z =0
\]
and since $\psi_z\neq 0$ one obtains from the first two equations
\[
\varphi(x,y,z) = \varphi_1(x,y) + \varphi_2(z) \mbox{ and }
b = \beta \exp{(-\varphi_1)},
\]
while the third equation assumes the form 
\begin{equation}
\bigtriangleup\psi +\nabla\psi\cdot\nabla\varphi =0
\label{voltage}
\end{equation}
At this point the NM have been reduced to the system of just two scalar 
equations (\ref{last_st}) and (\ref{voltage}).
Denote $f_1 = \exp{\varphi_1}$ and $f_2 = \exp{\varphi_2}$ and 
assume in addition 
\[\psi =\psi(z)\]
so that
\[
\psi'(z) = \epsilon \exp{(-\varphi_2)} = \frac{\epsilon}{f_2}
\]
It now follows from (\ref{last_st}) and (\ref{voltage}) that the triplet 
\begin{equation}
\vec{B} = \frac{\beta}{f_1(x,y)}\frac{\partial}{\partial z},
\qquad
\vec{E} = \frac{\varepsilon}{f_2(z)}\frac{\partial}{\partial z},
\label{mag_and_el}
\end{equation}
and
\begin{equation}
f(x,y,z) = f_1(x,y)f_2(z)
\end{equation}
is a solution of the NM if only $f_1$ and $f_2$ satisfy a decoupled 
system of semi-linear elliptic equations
\begin{equation}
-f_2''(z) - \frac{\varepsilon ^2}{f_2(z)} = \nu_2f_2(z) 
\label{charge_stripe}
\end{equation}
\begin{equation}
-\bigtriangleup f_1(x,y) +\frac{\beta^2}{f_1(x,y)} =\nu_1f_1(x,y). 
\label{vortex}
\end{equation}

At this point, I would like to emphasize one more time that 
in a field theory one looks for {\em strong} solutions, i.e.
solutions that satisfy equations in the classical sense almost 
everywhere. Typically, such solutions are smooth except for singularities
supported on a union of closed submanifolds. Furthermore, geometrically
invariant derivatives of the resulting fields in the distributional
sense signify charges. With this understood, let us briefly turn  
attention to equation (\ref{charge_stripe}). One wants to avoid holding
the reader hostage to the formal analysis of this elementary equation
which might be somewhat distracting. Thus, I have chosen
to briefly describe the solutions qualitatively
leaving aside technical details that can be easily reconstructed aside
by the reader. First, one notes that if $\nu_2>0$ then a solution
is concave, while for $\nu_2<0$ it will be convex for large values
where $f_2^2>-\varepsilon ^2/\nu_2$. Assuming formally that
$f_2$ is a function of $f_2'$ (piecewise), one reduces
(\ref{charge_stripe}) to the first order equation
\[
 \frac{df_2}{dz}=\pm\sqrt{c-\nu_2f_2^2-\varepsilon ^2\ln{f_2^2}}.
\] 
Thus, there are essentially two types of positive solutions,
depending on the actual values of constants $c,\varepsilon ,\nu_2$.
The first type includes solutions that assume value $0$ at a certain point
$z_0$ and increase monotonously to infinity as $z\rightarrow\infty$
as well as the symmetric solutions defined between $-\infty$ and some 
point, say $z_0$ again, where they reach $0$. These solutions require
$\nu _2 <0$ and they asymptotically look like $\exp{(\pm(-\nu_2)^{1/2}z)}$
One can use both branches in order to put together a strong solution that
forms a cusp or a jump discontinuity at $z_0$.
The second type consists of solutions that are concave, rise to the highest
peak at $f_2=m$, when $c-\nu_2f_2^2-\varepsilon ^2\ln{f_2^2}=0$,
and fall off to $0$ on both sides in finite time while
being differentiable in-between.
Selecting the constants and combining both
types of solutions piecewise segment-by-segment one obtains
strong solutions $f_2$ that in turn provide electric fields according
to formula (\ref{mag_and_el}).

Since, with the exception of the trivial
constant solution, there are no global smooth solutions, one concludes
that either $\vec{E}$ is constant or there exist charge stripes located
at planes $z=\mbox{const}$ where $f_2(z)$ has singularities.
The distributional derivative is in each case equal to the Dirac measure  
concentrated at $z=\mbox{const}$ as above and scaled by the size of 
the jump, and classical derivatives on both sides of the singularity. Even in
absence of a jump, the charge will switch from negative to positive
thus forming what can be amenably called a charge-stripe. An example
of this is shown in {\em Fig.1}.

It is much more difficult to figure out solutions of the second equation. 
I refer the reader to \cite{sowa3} for a more thorough analysis,
while here I will just briefly summarize my previous findings.
Solutions of equation (\ref{vortex}) correspond to critical points 
of the functional  
\[
L(f_1)= \frac{\frac{1}{2}\int |\nabla f_1|^2 + \beta^2\int
\ln(f_1)}{\int f_1^2}  
\]
which is neither bounded below nor above, so that one is looking at
the problem of existence of {\em local} extrema.
The equation always admits a trivial constant solution.
But, as it is shown in \cite{sowa3}, it also possesses
nontrivial vortex lattice solutions. More precisely, if $\beta$ is 
larger than a certain critical value then there is a nonconstant 
doubly periodic function $f$ which satisfies the finite difference version 
of (\ref{vortex}) everywhere except at a periodic lattice of isolated 
points, one point per each cell. 
In this way, a lattice of flux tubes, cf. {\em Fig.2},
emerges as a solution of the NM.
For the time being, the proof 
of this fact relies on finite-dimensionality essentially,
and does not admit a direct generalization to the continuous-domain case. 
However, physical parameters, like $\int f^2$ and $\beta$, 
are asymptotically independent of the density of discretization. 
Thus, I conjecture existence of the
continuous domain solutions that satisfy the equation a.e. in the 
classical sense and retain the particular vortex morphology.
Presently, the essential obstacle to proving this conjecture 
is lack of a regularity theory for the discrete vortex solutions.
The proof in \cite{sowa3} is carried out in the (discretized) torus
setting. One believes that vortex type solutions exist on any closed 
(orientable) surface. 

\section{Topological Quantum Numbers}
Every gauge theory comes equipped with an associated set of topological
invariants---usually characteristic classes of the bundles used to introduce the 
gauge field.  Articles \cite{Laugh}, \cite{LaughA}, \cite{LaughB}
teach us how such topological invariants
may be manifested in an electronic system as observable quantum numbers. 
The Nonlinear Maxwell Theory is naturally equipped with two kinds of topological
invariants. On one hand,
one has the first Chern class of the original $U(1)$-bundle.
Additionally, we will see below that in the case of static solutions the NM give us 
an additional set of  invariants defined directly by the foliated structure of 
the underlying three-manifold.
(In the discussion below, I generally assume for the sake of simplicity
that $M$ is a closed orientable manifold unless stated otherwise.)
In this section I will make an effort only to identify rather than exploit to
the fullest the geometric and topological ramifications of this nonlinear theory of 
Electromagnetism. To gain some initial impetus, let us be guided by
the following question
\begin{center}
\parbox{13cm}{\em{
What are the necessary and sufficient conditions on a 
Riemannian three-manifold $M$ for the NM to admit a separation 
of variables of the type seen in the previous section, i.e.
for the equation (\ref{vortex}) to decouple so that  
its solutions will generate magnetic-flux-tube type solutions 
on $M$?}
}
\end{center}
A question of this type is typical in algebraic topology where one is asking
about global obstructions to the presence of certain algebraic factorization 
properties of analytic objects, like linear differential equations as it is the case
for, say, the de-Rham
cohomology groups. In our case, the equations are nonlinear, but the
principle remains the same. The importance of these questions for practical 
issues of Electromagnetism is twofold. First, one wants to know how big is the set 
of possible configurations---especially in the absence of the superposition principle.
Secondly, I believe
the topological invariants displayed below are directly on target in an
effort to
explain and describe the nature of certain rigid structures,
like the Quantum Hall Effects, that physically occur in electronic systems. 

First, it needs to be emphasized that the static field equations
I want to consider, i.e. the equations that descend from the
four-dimensional spacetime via time-freezing coefficients, are distinct
from the equations
(\ref{syst0}-\ref{syst2}) considered directly on a three manifold.
Secondly, the equations
(\ref{first_st}-\ref{last_st}) are only valid on a Euclidean space. 
The geometry behind these equations is easier to identify when they 
are rewritten in an invariant form that can be considered on
any three-manifold in a coordinate independent setting.

Fix a Riemannian metric on $M$ with scalar product $<.,.>$ extended to include
measuring differential forms.
Denote by $B$ and $E$ the forms dual
to the magnetic and electric field vectors; recall notation $\varphi = \ln{f}$
and put $a=1$. The static NM assume the form
\begin{equation}
d E = 0
\label{d}
\end{equation}
\begin{equation}
\delta B = 0
\label{delta}
\end{equation}
\begin{equation}
\star (\star dB\wedge E) + (\delta E)B = -<E, B>d\varphi
\label{stard}
\end{equation}
\begin{equation}
d B\wedge B = 0
\label{dwedge}
\end{equation}
\begin{equation}
\bigtriangleup \varphi + |d\varphi |^2 + |E|^2 -|B|^2 + \nu =0.
\label{phi_long}
\end{equation}
Equation (\ref{dwedge}) is the familiar Frobenious condition on integrability
of the distribution of planes given by $\ker B$. One always assumes
$B$ is nonsingular a.e. so that the distribution is {\em a priori} also
defined a.e. For convenience, it is assumed throughout this section that the
foliation determined by $\ker B$ is smooth. 
(It is quite clear that for the flux-tube type solutions the
distribution extends through the singular points and is defined everywhere.
At this stage, however, it is hard to make a formal argument to this
effect, hence the {\em a priori} assumption.) 
The condition of smoothness implies that the
three-manifold $M$ must have vanishing Euler characteristic.
In particular,
singular foliations, some of which may be associated with other types
of solutions of the NM, are excluded from the discussion below.

It follows that there is a $1$-form $\alpha$,
known as the Godbillon-Vey form, such that 
\[dB=\alpha\wedge B.
\]
This form is not defined uniquely. However, as is well known, $d(\alpha\wedge d\alpha )=0$
and the Godbillon-Vey (GV) cohomology class
\[ 
[\alpha\wedge d\alpha ]_{H^3(M)}
\]
is uniquely defined. On a three manifold this class can be evaluated 
by integration resulting in a Godbillon-Vey number
\[
 Q = \int_{M}\alpha \wedge d\alpha . 
\]
This invariant poses many interesting questions that have not been fully resolved 
by geometers yet. Below, I will justify two observations. 
First, the condition of existence of the magnetic 
flux-tube solutions imposes both local and global restrictions on  the foliation. Second,
magnetic flux-tube solutions exist in topologically nontrivial situations with nonzero 
GV-number $Q$. This is formally summarized in the two propositions that follow. 
They are far from the most general statements that can be anticipated in this direction,
but are also nontrivial enough to suggest a conjecture regarding {\em quantization} of the 
GV-number that I will formulate following Proposition \ref{P1}.

Consider {\em a priori} a foliation given by $\ker B$ locally. 
First, one introduces a local coordinate patch $(x,y,z)$ such that
the foliation is given
by the $(x,y)$-planes  and $|dz|=1$. In particular 
\[
B=\beta (x,y,z)dz.
\]
 Let $\gamma = g(x,y,z) dx\wedge dy$ denote the volume element on a leaf.
 One has
$\star B = \beta (x,y,z)\gamma$. Equation (\ref{delta}) becomes 
$d(\beta (x,y,z)g(x,y,z)dx\wedge dy)=0$. Thus, there is a function $\chi =\chi (x,y)$
such that
\[ 
\beta (x,y,z) = \frac{\chi (x,y)}{g(x,y,z)}.
\]
A calculation analogous to that in the previous section shows that the whole system 
(\ref{delta}-\ref{phi_long}) is reduced to 
\begin{equation}
\label{logs}
\left(\ln{\frac{\chi (x,y)}{g(x,y,z)}}\right)_x = -\varphi _x, \qquad
\left(\ln{\frac{\chi (x,y)}{g(x,y,z)}}\right)_y = -\varphi _y
\end{equation}
\begin{equation}
\label{delE}
\delta E +<E,d\varphi > =0
\end{equation}
\begin{equation}
\label{philap}
\bigtriangleup \varphi + |d\varphi |^2 + |E|^2 -\left(\frac{\chi (x,y)}{g(x,y,z)} \right)^2 + \nu =0.
\end{equation}
Observe that in order to obtain factorization 
\begin{equation}
\label{split}
\varphi(x,y,z) = \varphi_1(x,y) +\varphi_2(z)
\end{equation} 
it is necessary and sufficient that 
\begin{equation}
\label{zindep}
g=g(x,y),
\end{equation}
i.e. a priori dependence of $g$ on $z$ is dropped. 
If that holds, the equations (\ref{delE}) and (\ref{philap}) can be 
decoupled with an additional Ansatz $E=e(z)dz$. One also has 
that $\chi /g = b\exp{(-\varphi_1)}$ for a constant $b$
and 
\begin{equation}
\label{flux}
\bigtriangleup_{x,y} \varphi_1 + |d\varphi_1 |^2 -b^2 \exp{(-2\varphi_1)}+ \nu =0.
\end{equation}
Conversely, if (\ref{flux}) and (\ref{split}) hold, then so must 
(\ref{zindep}) and the mean curvature $h$ of a leaf vanishes. 
Indeed, by definition
\[
h = \delta \left(\frac{1}{|B|}B\right) = -\star  d (g(x,y)dx\wedge dy) =0.
\]
This implies
\begin{prop}
\label{P1}
For the existence of flux-tube type solutions---in the sense of existence of
factorization (\ref{split}) and decoupling of equation (\ref{flux})---it is necessary that
the foliation given by $\ker{B}$ be taut, i.e. the mean curvature of leaves
must vanish. In particular $\pi_1(M)$ must be infinite.
\end{prop}
{\em Proof.} The first part has been shown above. 
The second part follows from a result of D. Sullivan \cite{Sul}
that he deduced from the result of Novikov on the existence of
a closed leaf that is a torus
(cf. \cite{Nov}, and \cite{Tond} for additional
general material and references).$\Box$

In particular, there are no flux-tube type solutions of the NM 
that would conform with the Reeb foliation \cite{Reeb}. 
This is a practical issue since the Reeb foliation exists on 
a solid torus, so that in principle it might be observed experimentally
which would be inconsistent with the theory at hand.
This fact is also interesting for another reason. Namely, according to the
celebrated theorem by Thurston in \cite{Thurs} each real number
may be realized as the Godbillon-Vey number for a certain codimension one
foliation on the three-sphere $S^3$. The known proof of this result uses
the Reeb foliation in an essential way. I do not know if this fact is
canonical,
i.e. if the presence of the Reeb foliation is necessary for the result to
hold,
but if it turns out to be so then excluding the Reeb foliation from the game
should result in a reduction of the range of the G-V number, possibly to
a discrete subset of the real line. In such a case, the resulting set of
the G-V numbers accompanying flux-tube type solutions of the NM would also
be discrete. This is consistent with my expectation
that these invariants must be related with
(both the integer and the fractional) Quantum Hall Effects.
Future research should bring a resolution of this problem.

Another observation is that the factorization given by (\ref{split}) and 
(\ref{flux}) does exist in topologically nontrivial situations. More
precisely, I want to consider solutions of the NM on 
$PSL(2,R)$ and its compact factors. These three-manifolds are equipped 
with interesting codimension one foliations known as the Roussarie 
foliation \cite{Rouss}. Let the Lie algebra $\sc{sl}(2,R)$ be given by
\[
[X, Y^+] = Y^+,\quad [X,Y^-]=-Y^-,\quad [Y^+,Y^-]=2X.
\] 
Pick a metric on $PSL(2,R)$ in which the corresponding 
left-invariant vector fields $X$, $Y^+$, and $Y^-$ are orthonormal
and let $\mu $, $\nu^+$, and $\nu^-$ be the corresponding dual 
$1$-forms. One checks directly that
\[
 d\nu^- = \mu \wedge \nu^-.
\]
so that the distribution $\ker{\nu^-}$ is integrable and $\mu$ is the 
GV-form of the resulting foliation. In particular, one can introduce local 
coordinates $(x,y,z)$ such that $\partial_x = X$, $\partial_y = Y^+$,
$\partial _z = Y^-$. This foliation descends
to compact factors of $PSL(2,R)$ that can each be identified with 
$T_1M_g$---the total space of the unit tangent bundle of the 
hyperbolic Riemann surface 
of genus $g$ that depends on our choice of the co-compact subgroup
acting on $PSL(2,R)$ by isometries. Moreover, the GV-integrand 
$\mu\wedge d\mu$ is proportional to the natural volume form on the 
three-manifold. As a result of this, the corresponding GV-numbers
\[
 Q = \int_{T_1M_g} \mu\wedge d\mu = -2Vol(M_g)
\] 
assume values in a discrete set. I want to look for solutions of the
NM that satisfy the Ansatz 
\begin{equation}
 B = \beta \nu^-. 
\end{equation}
In particular, (the Frobenious) equation (\ref{dwedge}) is satisfied 
automatically. Moreover, since 
$\star dB = (Y^-\beta)\mu\wedge\nu^+\wedge\nu^-$, equation 
(\ref{delta}) implies
\begin{equation}
\label{betaind}
\beta = \beta (x,y).
\end{equation}
As before, one checks that (\ref{stard}) implies (\ref{delE})
as well as
$X\varphi = - X\ln\beta$ and $Y^+\varphi = - Y^+\ln\beta$.
In consequence, one again has (\ref{split}) and assuming $E=e(z)\nu^-$
as before one obtains (\ref{flux}). In consequence, the following holds true.
\begin{prop}
The Roussarie folitions on $PSL(2,R)$ and its compact factors
satisfy the factorization condition for the existence of magnetic 
flux-tube type solutions in the sense that the tangent distribution 
can be expressed as $\ker{B}$ a.e. and one can reduce the NM 
to the form (\ref{split}-\ref{flux}).
 \end{prop}
In a similar way one can obtain factorization (\ref{split}) and 
(\ref{flux}) for other foliations, like the natural foliation 
on say $S^2\times S^1$.

\section{More on the Physical Framework of the NM}

It is natural to ask if the NM descend from a Lagrangian
functional depending on the two variables $A$ and $f$,
say $\Phi(A,f)$, via the Euler-Lagrange calculus of variations.
The answer is negative as one can easily see considering
that in general a gradient must pass the second derivative test: 
\[
\frac{\delta ^2}{\delta A\delta f}\Phi = 
 \frac{\delta ^2}{\delta f\delta A}\Phi
\]
---a condition that cannot be satisfied by the expressions
in (\ref{syst0}-\ref{syst2}) viewed as the gradient, say
$(\frac{\delta}{\delta A}\Phi , \frac{\delta}{\delta f}\Phi)$,
of an unknown functional $\Phi$.
This suggests that the NM may constitute just a part of a broader
theory that would encompass additional physical fields. In other words,
the equations (\ref{syst0}-\ref{syst2}) would have to be coupled to
some other equations via additional fields. In addition, such coupling would
have to induce only a very small perturbation of the present picture that
one believes is essentially accurate.
Such possibilities may become more accessible in the future.
Among other, perhaps related goals is that of deriving the NM equation
directly from the microscopic principles.

The well-known analogy between the Quantum Hall Effects and 
High-$T_c$ Superconductivity suggests that there should exist
vortex lattices involving the so-called {\em filling factor} 
(microlocally a constant scalar) that plays a major role in the 
description of {\em Composite Particles}. The NM describe exactly 
this type of a vortex-lattice. 
Simulation and theory show that this system conforms with the 
experimentally observed physical facts. It stretches the domain 
of applicability of the Maxwell theory to encompass phenomena 
such as the {\em Magnetic Oscillations}, {\em Magnetic Vortices}, 
{\em Charge Stripes} that occur in low-temperature electronic 
systems exposed to high magnetic fields.  

There are other systems of PDE that admit vortex-lattice solutions 
and are conceptually connected with Electromagnetism, like
the well known Ginzburg-Landau equations valid within the framework of
low $T_c$ type-II superconductivity, or
the Chern-Simons extension of these equations which, some researchers have 
suggested, may be more relevant to 
the Fractional Quantum Hall Effect and/or High-$T_c$ Superconductivity,
cf. \cite{Zhang}.
The free variables of these equations are the so-called 
{\em order parameter} (a section of a complex line-bundle) and a 
$U(1)$-principal connection, both of them containing topological 
information. In the case of NM, all the topological information is 
contained in one of the variables, i.e. the principal connection, while 
the other is a scalar function.
An additional advantage of the NM is in that it remains meaningful in 
three-plus-one dimensions just as well as in the two-dimensional setting.
I would also like to mention that recently other researchers have introduced 
nonlinear Maxwell equations of another type in the context of the Quantum Hall 
Effects, cf. \cite{Frohlich}. The NM theory presented in this and the preceding 
articles of mine is of a different nature. Finally, although this is far from 
my areas of expertise and the remark should be received as completely {\em ad hoc}, 
I would  also like to mention 
that yet another context in which foliations come in touch with the 
Quantum Hall Effect is that of noncommutative geometry, cf. \cite{Connes}.

Let me conclude with a question that may suggest yet another point of view. 
Namely, is there a coalescence between the nonlinear PDEs (in the form of 
the NM) and the (Quantum) Information Theory? 
As it was pointed out, construction of error correcting 
codes may unavoidably require manipulating quantum information 
at the topological level. Anyhow, this is how I have understood the essential 
thought in \cite{Freedman}. Adopting this paradigm would strongly suggest that 
the effective language of quantum computation should be costructed at many levels, 
including that of the mesoscopic field theory in parallel with the language derrived
from the basic principles as it is done now. Future research will likely 
better clarify these issues.

\newpage
\noindent
{\em Fig.1} An example of a strong solution of (\ref{charge_stripe}).
$f=f_2(z)$ is a positive function, the electric field is given by formula
(\ref{mag_and_el}). The resulting charge distribution is obtained by
evaluating $\nabla\cdot\vec{E}$. (In general, 
$\nabla\cdot\vec{E}$ is understood in the distributional sense).
Charge is concentrated along certain plains $z=\mbox{const}$. This is
the basic appearance of charge stripes---intertwining concentrations
of positive and negative charges. (One should compare
this static picture with the description of moving charge stripes
in section \ref{charge_wave}.)

\vspace{.1cm} \vbox spread 3in{}
\includegraphics{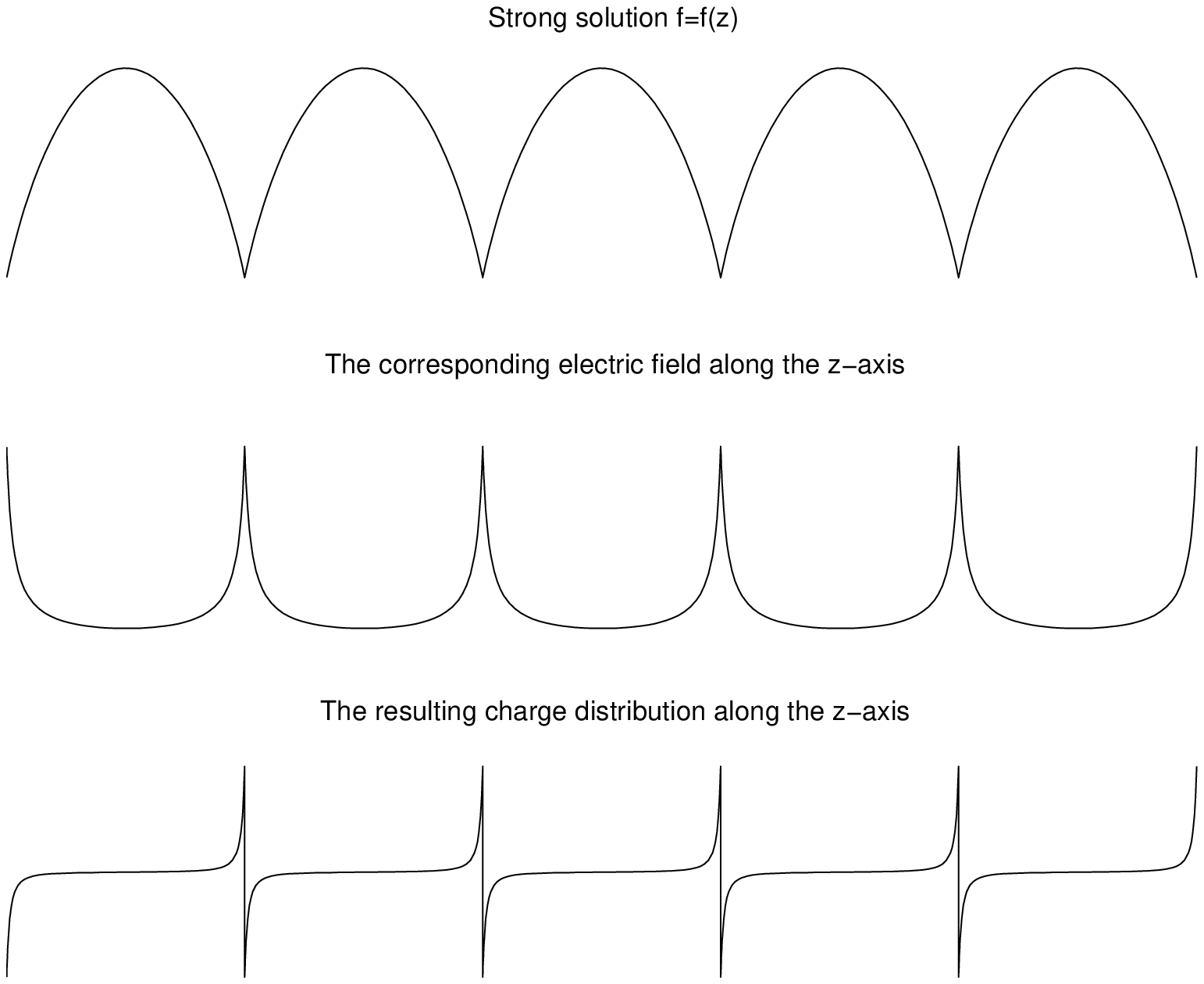}
\newpage
\noindent
{\em Fig.2} The luminance graph of $f=f_1(x,y)$ that solves (\ref{vortex}).
The corresponding magnetic field on the right is obtained via
(\ref{mag_and_el}).

\vspace{.1cm} \vbox spread 3in{}
\includegraphics{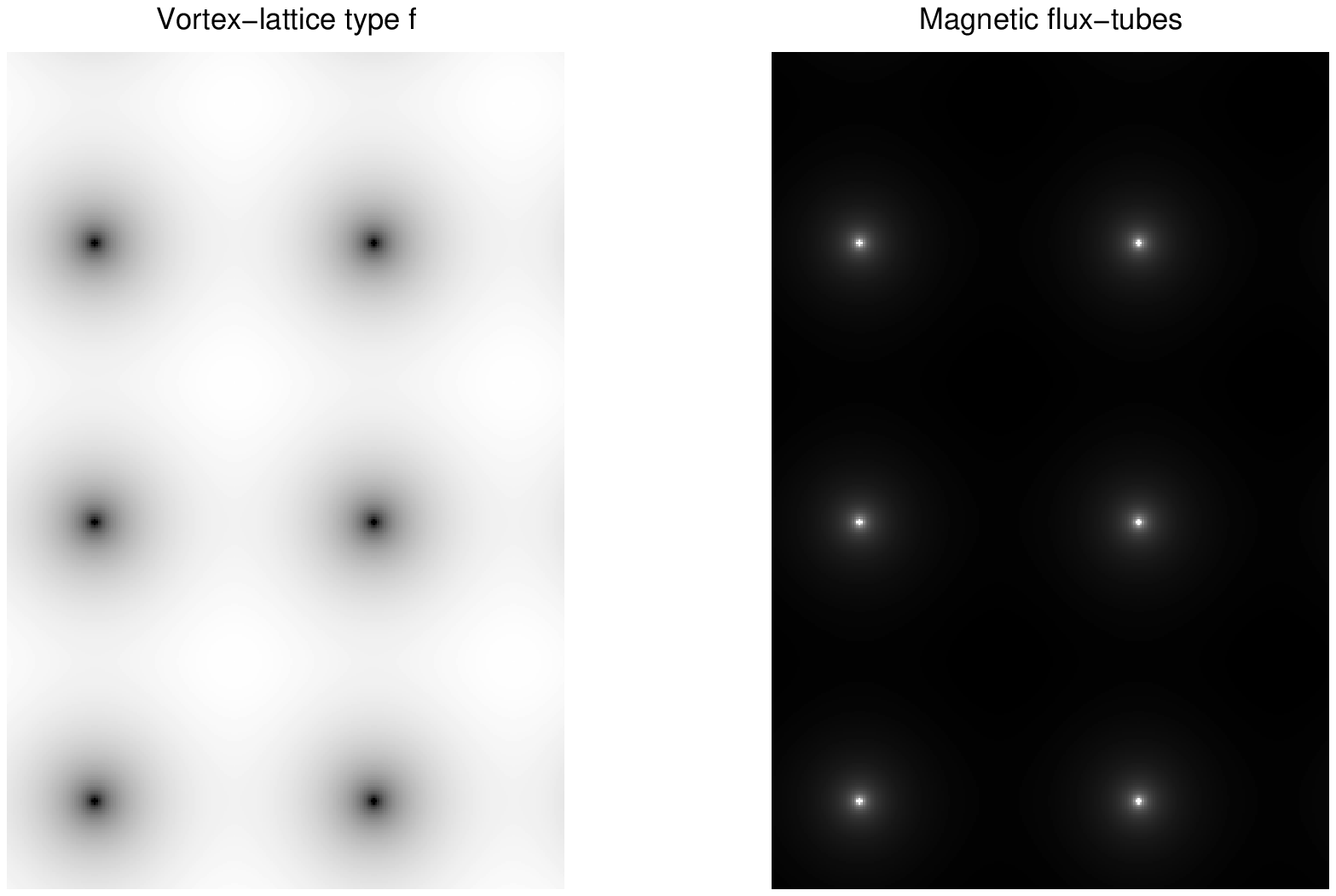}
\end{document}